\documentclass[pra,aps,twocolumn,superscriptaddress,nofootinbib,balancelastpage]{revtex4-1}
\usepackage{latexsym,amssymb,bm,bbold,amsfonts,amstext,graphicx,bbm,bm,relsize,dsfont,times}
\usepackage[dvipsnames]{xcolor}
\usepackage{enumerate}

\usepackage{amsmath} 
\usepackage{amsthm} 
\usepackage[unicode=true, breaklinks=false, pdfborder={0 0 1}, backref=false, colorlinks=true, linkcolor=blue, citecolor=blue]{hyperref}
\usepackage{subfiles}

\usepackage{tcolorbox}

\usepackage{tikz}

\begin{document}

\def\be{\begin{equation}}
\def\ee{\end{equation}}
\def\bea{\begin{eqnarray}}
\def\eea{\end{eqnarray}}

\title{Exact nonequilibrium quantum observable statistics: A large-deviation approach}

\author{Stefano Gherardini}
\email{gherardini@lens.unifi.it}
\affiliation{\mbox{Department of Physics and Astronomy \& LENS, University of Florence,} via G. Sansone 1, I-50019 Sesto Fiorentino, Italy.}
\affiliation{\mbox{INFN Sezione di Firenze}, via G. Sansone 1, I-50019 Sesto Fiorentino, Italy.}

\begin{abstract}
The exact statistics of an arbitrary quantum observable is analytically obtained. Due to the probabilistic nature of a sequence of intermediate measurements and stochastic fluctuations induced by the interaction with the environment, the measurement outcomes at the end of the system's evolution are random variables. Here, we provide the exact large-deviation form of their probability distribution, which is given by an exponentially decaying profile in the number of measurements. The most probable distribution of the measurement outcomes in a single realization of the system transformation is then derived, thus achieving predictions beyond the expectation value. The theoretical results are confirmed by numerical simulations of an experimentally reproducible two-level system with stochastic Hamiltonian.
\begin{description}
\item[PACS numbers]05.30.-d, 03.65.Yz, 03.65.Ta
\end{description}
\end{abstract}

\maketitle

\section{Introduction}

Observables of a quantum system out-of-equilibrium return random outcomes fluctuating with a specific probability distribution \cite{BookJacobs}.
By following the operational approach \,\cite{DaviesBook}, a quantum stochastic process\,\cite{BookPetruccione,ReviewHuelga} can be envisaged as a nonequilibrium transformation modeled as the composition of stochastic evolutions characterized by semi-classical fluctuations of the system parameters\,\cite{GherardiniErgodicity,MuellerAnnalen,RossiPRA2017} and intermediate quantum measurements \cite{CampisiPRL2010,ThielPRL2018}. In the former case, especially in quantum computing, stochastic fluctuations can arise from imperfections in the physical realization of the quantum system or through the interaction with the environment\,\cite{MishraNATCOMM2018}.
On the other hand, repeated quantum measurements could correspond to a process exchanging photons with the environment\,\cite{HekkingPRL2013} or could be adopted to ensure the protection of coherent evolutions of a quantum system by decoherence (quantum Zeno dynamics)\,\cite{FacchiPRL2002}. Experimentally, together with strong coupling methods, such dynamics have been realized in several physical setups such as solid-state spins, superconducting qubits or ultracold atomic gases\,\cite{UhrigPRL2007,LangeScience2010,PokharelPRL2018,SchaferNAT2014,SignolesNAT2014,GherardiniErgodicity}. They are also relevant in quantum metrology\,\cite{PezzeReview2018} to probe the phase evolution of an atomic ensemble by means of interleaved interrogations and feedback corrections\,\cite{KohlhaasPRX2015,SchioppoNAT2017}. In this scenario, by observing only once the dynamics of the system (single realization), the ensemble average of the measurement outcomes does not provide complete information about the statistics of the measured results. This becomes more evident when also one rare random event, i.e.\,a stochastic fluctuation with very small probability, occurs within the system dynamics\,\cite{Chantasri2013,WeberNature2014,Chantasri2015}. Such a concept is at the heart of the Large Deviation (LD) theory \cite{Ellis1,Touchette1}, dealing with the exponential decay of probabilities associated to large fluctuations in stochastic classical \cite{Dembo1} and quantum systems\,\cite{Gallavotti2002,Netocny2004,GarrahanPRL2010,LesanovskyPRL2013,PigeonPRA2015,Collura2019}.

In this paper, we derive a closed-form expression for the outcomes'statistics obtained by measuring a quantum system repeatedly monitored by an external observer and whose dynamical evolution depends on random parameters. In particular, we prove that for a sufficiently large number $m$ of intermediate quantum measurements, the probability distribution of the last measurement results obeys the so-called \emph{LD principle}\,\cite{Touchette1,GherardiniNJP,thesis_Stefano}. This means that the behaviour of the measurement outcomes' distribution is a decaying exponential in $m$, whose exponent is equal to the relative Shannon entropy between the configurations of the stochastic system dynamics. In other words, only a rigorous description of the occurrence combinatorics of the parameters defining the stochastic evolution of the system allows for the full characterization of the outcomes' statistics, also \emph{beyond} the Gaussian approximation given by the sole description of the measurement apparatus.

\section{Model}
\label{model}

Let us consider an arbitrary quantum system $\mathcal{S}$ within the Hilbert space $\mathcal{H}$. We assume that $\mathcal{S}$ is initialized in the quantum state with density matrix $\rho_0$ and that the Hamiltonian $H$ of the system is time-independent. At the level of the single trajectory, we assume that the random interaction between $\mathcal{S}$ and an environment $\mathcal{E}$ gives rise to a sequence of stochastic dynamical evolutions, separated by consecutive quantum projective measurements\,\cite{footnote1}, following the postulates of quantum mechanics\,\cite{Sakurai1994}. Hereafter, the index $j$ denotes the dimension of $\mathcal{H}$, while the index $\alpha$ denotes the instants composing the temporal sequence of measurements. More specifically, we assume that the first $m-1$ measurements are performed on the quantum observable $\mathcal{O} \equiv \sum_{j}o_{j}\Pi_{o_j}$, where $o_j$ are the outcomes of $\mathcal{O}$ and $\{\Pi_{o_j}\}$ is the set of projectors corresponding to the measured eigenvalues at time $t_{\alpha}$. The $m-$th measurement, instead, is performed on the quantum observable $\Theta \equiv \sum_{j}\theta_{j}\Pi_{\theta_{j}}$, whose outcomes $\theta_j$ are recorded by the observer. According to the postulate of quantum measurement, the state $\rho_{\alpha}$ of $\mathcal{S}$ after the projective measurement at $t_{\alpha}$ is identically equal to one of the projectors defining the measurement observable. Then, between each projection event the system undergoes a dynamics, that is governed by the Hamiltonian $H$ and described by the completely-positive and trace-preserving quantum map $\Phi(t_\alpha,t_0)[\rho_0] \equiv \Phi_{\alpha}[\rho_0]$ \cite{Caruso_RevModPhys_2014}. If we assume quite large number of intermediate quantum measurements, the dynamics between measurements is described by a unitary operator, so that $\Phi_{\alpha}[\rho_{\alpha-1}]$ is simply given by the super-operator $\mathcal{U}_{\alpha}[\rho_{\alpha-1}] \equiv U_{\alpha}\rho_{\alpha-1}U_{\alpha}^{\dagger}$, where $U_\alpha \equiv \exp(-iH\tau_\alpha)$, $\hbar$ is set to unity and $\tau_{\alpha} \equiv t_{\alpha} - t_{\alpha - 1}$. We also assume that, in accordance with the recently introduced stochastic quantum Zeno phenomena \cite{GherardiniNJP,GherardiniErgodicity,MuellerAnnalen}, there exists for each propagator $U_{\alpha}$ at least one dynamical parameter $\lambda$, that is a fluctuating variable. For example, one could consider as in \cite{Gherardini_heat_stat,MukherjeePRE2018} that each $\lambda_\alpha$ is equal to the time interval $\tau_{\alpha}$, with $\tau_\alpha$ random. Moreover, the $\lambda$'s between the measurements are taken as constant and have a different random value only after the occurrence of a new measurement, according to the probability density function $p(\lambda)$. Here, $\vec{\lambda} \equiv (\lambda_1,\ldots,\lambda_{m})$ denotes the sequence of dynamical parameters $\lambda_{\alpha}$ in a single realization of the system transformation. The fluctuating dynamical parameters $\lambda_{\alpha}$, then, are taken as independent and identically distributed (i.i.d.)\,random variables sampled from $p(\lambda)$, since the environment is a-priori unknown.

The stochastic nature of the measurement outcomes $\theta_{m}$'s at time $t_{m}$ lies in the specific values assumed by $\vec{o}$ and $\vec{\lambda}$. Thus, being the dynamics of $\mathcal{S}$ stochastic, the single realization of the system density matrix $\rho_{m,\vec{o},\vec{\lambda}}$ at the end of its evolution is a fluctuating variable, i.e., given the sequences $\vec{o}$ and $\vec{\lambda}$, it is mapped into
\begin{equation}
\rho_{m,\vec{o},\vec{\lambda}} = \frac{\mathcal{P}_{\theta_{m}}\mathcal{U}_{m}\mathcal{P}_{o_{m-1}}\mathcal{U}_{m-1}\cdots\mathcal{P}_{o_1}\mathcal{U}_{1}[\rho_0]}{p_{\theta_{m}}(\vec{o},\vec{\lambda})},
\end{equation}
where $\mathcal{P}_{\mu_{\alpha}}[(\cdot)] \equiv \Pi_{\mu_{\alpha}}(\cdot)\Pi_{\mu_{\alpha}}$ is the measurement super-operator acting on $(\cdot)$ at $t_{\alpha}$, with $\mu\in\{o,\theta\}$. Consequently
\begin{equation}\label{prob_outcomes}
p_{\theta_{m}}(\vec{o},\vec{\lambda}) \equiv \textrm{Tr}\left[
\mathcal{P}_{\theta_{m}}\mathcal{U}_{m}\mathcal{P}_{o_{m-1}}\mathcal{U}_{m-1}\cdots\mathcal{P}_{o_1}\mathcal{U}_{1}[\rho_0]
\right],
\end{equation}
i.e.\,\,$p_{\theta_{m}}(\vec{o},\vec{\lambda}) = \textrm{Tr}[\Pi_{\theta_{m}}\rho_{m,\vec{o},\vec{\lambda}}]$, denotes the conditional probability to obtain the outcome $\theta_{m}$ from the measurement of $\Theta$ given the specific realization of the sequences $\vec{o}$ and $\vec{\lambda}$.

\section{Quantum observable statistics}
\label{q_obs_stat}

\begin{figure}[t!]
  \includegraphics[scale = 0.605]{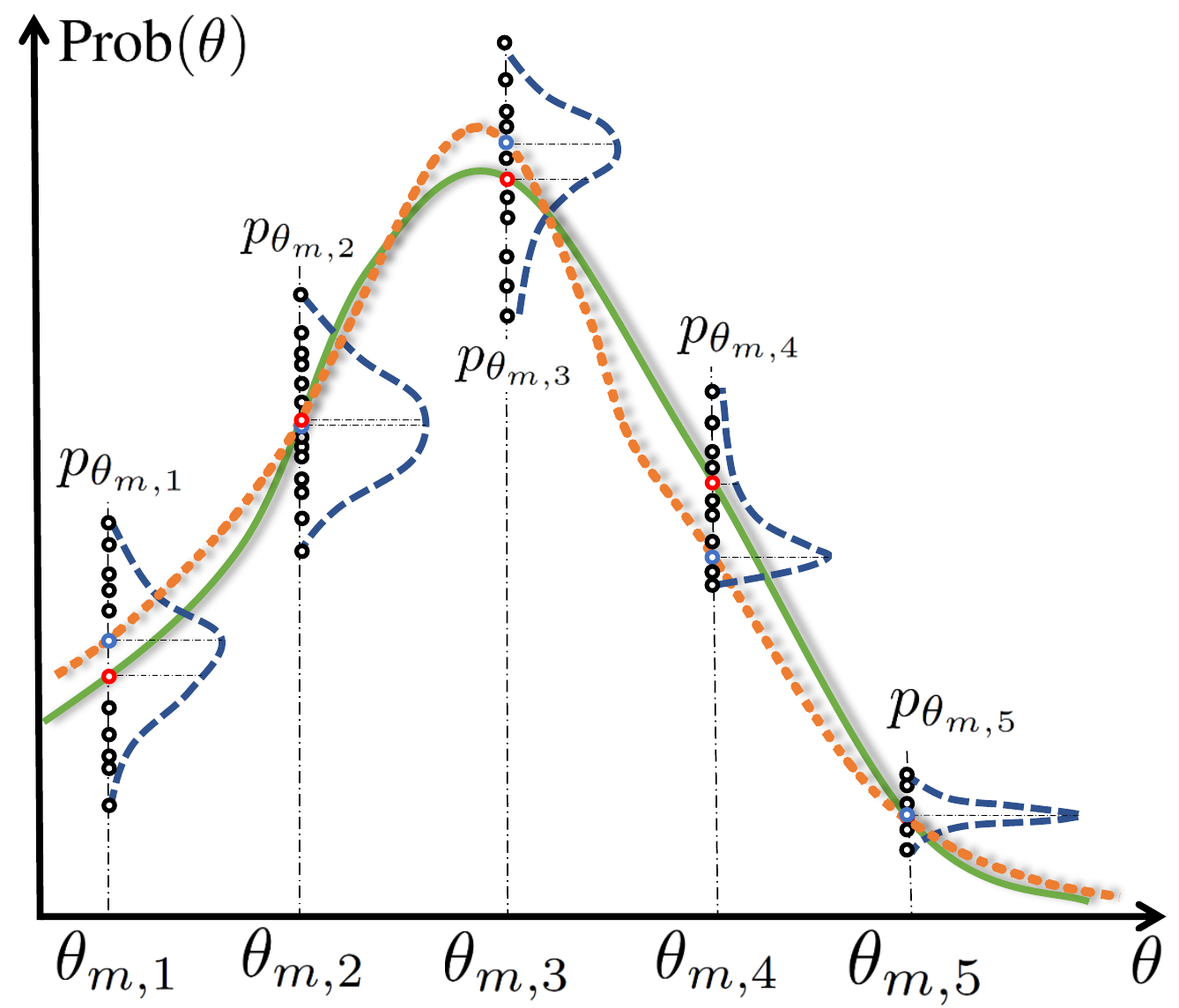}
  \caption{Statistics of $\Theta$'s outcomes. By repeating several times the stochastic evolution of the system and measuring each of the outcomes $\theta_{m,j}$ at the final time instant $t_{m}$, an ensemble of conditional probabilities $p_{\theta_{m,j}}$ is obtained, where each of them is computed after the single realization of the system dynamics. Thus, just by counting the occurrence relative frequencies of the $p_{\theta_{m,j}}$'s, one can derive the corresponding probability distributions (blue dashed lines). If the number of realizations is relatively small, such distributions do not obey the Gaussian approximation. This means that one has to distinguish between two different statistics for $\theta$'s: One (green solid line) linking the ensemble averages $\langle p_{\theta_{m,j}}(\vec{o},\vec{\lambda})\rangle$ (red dots) of the conditional probabilities for each measurement outcome, and the other (orange dotted line) -- also called \emph{most probable distribution} -- connecting all the realizations of $p_{\theta_{m,j}}$ in correspondence to the maximum value of the conditional probability distributions (blue dots).}
  \label{fig:fig1}
\end{figure}

In a single realization of the system transformation, $\rho_{m,\vec{o},\vec{\lambda}}$ and $\theta_{m}$ depend on $\vec{o}$ and $\vec{\lambda}$ and are random quantities. Thus, the following question naturally emerges: Which is the \emph{best} description for the statistics of the measurement results $\theta_{m,j}$ from the observation of $\mathcal{S}$ at $t_{m}$? Three possible answers, characterized by an increasing degree of prediction accuracy, can be provided. First, one could describe the probabilistic expected result from the measured outcomes by using the expectation value $\textrm{Tr}[\Theta\rho_{m}]$, with
\begin{equation}
\rho_{m} \equiv \langle\rho_{m,\vec{o},\vec{\lambda}}\rangle = \sum_{\vec{o}}\int_{\vec{\lambda}}d^{m}\vec{\lambda}\,p(\vec{\lambda})\rho_{m,\vec{o},\vec{\lambda}}
\end{equation}
and $p(\vec{\lambda})$ denoting the occurrence joint probability of the dynamical parameters $\lambda_{\alpha}$. It is worth noting that $\textrm{Tr}[\Theta\rho_{m}]$ is equal to the ensemble average of all possible measurement outcomes. Second, the probability distribution of $\Theta$'s outcomes (green line in Fig.\,\ref{fig:fig1}) can be introduced:
\begin{equation}\label{prob_vartheta}
\textrm{Prob}(\theta) = \sum_{j}\delta(\theta - \theta_{m,j})\overline{p}_{\theta_{m,j}}\,,
\end{equation}
where $\overline{p}_{\theta_{m,j}} \equiv \langle p_{\theta_{m,j}}(\vec{o},\vec{\lambda})\rangle = \textrm{Tr}[\Pi_{\theta_{m,j}}\rho_{m}]$ and $\delta(\cdot)$ is the Kronecker delta. Eq.\,(\ref{prob_vartheta}) defines the statistics of $\Theta$, while $\overline{p}_{\theta_{m,j}}$ (red dots in Fig.\,\ref{fig:fig1}) is the probability to obtain on average the $j-$th outcome $\theta_{m,j}$. Otherwise, the third option that we are here proposing is to adopt the probability distribution of the conditional probabilities $p_{\theta_{m,j}}(\vec{o},\vec{\lambda})$, i.e.\,$\textrm{Prob}(p_{\theta_{m,j}}(\vec{o},\vec{\lambda}))$ (in Fig.\,\ref{fig:fig1}, the blue dashed lines for each outcome $\theta_{m,j}$), defined over all the possible realizations of the sequences $\vec{o}$ and $\vec{\lambda}$. Only in this way, by deriving $\textrm{Prob}(p_{\theta_{m,j}}(\vec{o},\vec{\lambda}))$, one will be able to get the most probable statistics (orange dotted line) for the quantum observable $\Theta$ connecting the blue dots in Fig.\,\ref{fig:fig1}.

We start by observing that the $p_{\theta_{m,j}}(\vec{o},\vec{\lambda})$'s can be written as the product of the probabilities $|\langle\pi_{\mu_{\alpha-1}}|U_{\alpha}(\lambda_{\alpha})|\pi_{\mu_{\alpha}}\rangle|^2$ -- also called dynamical transition probabilities -- that the quantum state moves from $|\pi_{\mu_{\alpha-1}}\rangle$ to $|\pi_{\mu_{\alpha}}\rangle$ via the propagator $U_{\alpha}(\lambda_{\alpha})$ (see the Appendix for more details). Specifically, $|\pi_{\mu_{\alpha}}\rangle$'s are the eigenvectors that define the measurement projectors $\Pi_{\mu_{\alpha}}$, with $\mu$ equal to $\theta$ or $o$, depending on whether the observable $\Theta$ or $\mathcal{O}$ is measured. Thus, one has that
\begin{eqnarray}
&\displaystyle{p_{\theta_{m,j}}(\vec{o},\vec{\lambda}) = \prod_{\alpha=1}^{m}|\langle\pi_{\mu_{\alpha-1}}|U_{\alpha}(\lambda_{\alpha})|\pi_{\mu_{\alpha}}\rangle|^{2},\,\,\,\text{i.e.,}}&\label{p_theta_jm} \\
&\displaystyle{\overline{p}_{\theta_{m,j}} = \sum_{\vec{o}}\prod_{\alpha = 1}^{m}\int_{\lambda_{\alpha}}d\lambda_{\alpha}p(\lambda_{\alpha})q_{\alpha}(\mu_{\alpha-1},\mu_{\alpha},\lambda_{\alpha}),}&\label{p_average}
\end{eqnarray}
where $q_{\alpha} \equiv |\langle\pi_{\mu_{\alpha-1}}|U_{\alpha}(\lambda_{\alpha})|\pi_{\mu_{\alpha}}\rangle|^2$, $\pi_{\mu_0} \equiv \pi_0$ (with $\rho_0 = |\pi_{0}\rangle\langle\pi_{0}|$), $\mu_{\alpha} \equiv o_{\alpha}$ for $\alpha = 1,\ldots,m-1$ and $\pi_{\mu_{m}} \equiv \pi_{\theta_{m,j}}$.

\section{Large deviation formalism}
\label{LD}

We now derive the main result of this paper, namely a closed-form expression of the $p_{\theta_{m,j}}(\vec{o},\vec{\lambda})$ distributions. Notice that if we just characterize the quantum observable $\Theta$ at $t_m$ and at the same time the hypotheses of the central limit theorem are respected, then the statistics of its outcomes is well represented by a Gaussian probability distribution. In fact, with this assumption, outliers in the outcomes statistics are simply classified as the result of a non-modeled experimental noise on the measurement apparatus. Thus, by increasing the number of realizations, the occurrence probability of the outliers naturally decreases and the Gaussian distribution well fits the data. However, this evidence is no longer valid if we assume that the statistics of $\theta_{m,j}$'s is given by an arbitrary stochastic transformation governing the dynamics of the system. In such a case, the configuration space defined by the occurrence of random events during its evolution becomes exponentially larger, but allows for the description of the outliers'statistics with occurrence probability greater than zero and not satisfying the Gaussian approximations. Therefore, we expect to increase our prediction power on the outcomes' distribution by statistically characterizing the trajectories of the system before the measurement of $\Theta$. This amounts to computing the statistics of the dynamical transition probabilities $q_{\alpha}(\mu_{\alpha-1},\mu_{\alpha},\lambda_{\alpha})$ as defined in Eq.\,(\ref{p_theta_jm}). Briefly, the procedure to derive the exact LD form of $\textrm{Prob}(p_{\theta_{m,j}}(\vec{o},\vec{\lambda}))$ is to take the logarithm of the conditional probabilities $p_{\theta_{m,j}}(\vec{o},\vec{\lambda})$, i.e.\,$l_{\theta_{m,j}}(\vec{o},\vec{\lambda}) \equiv \ln p_{\theta_{j,m}}(\vec{o},\vec{\lambda}) = \sum_{\alpha = 1}^{m} \ln q_{\alpha}(\mu_{\alpha-1},\mu_{\alpha},\lambda_\alpha)$, compute its distribution and then apply the contraction principle from LD theory\,\cite{Touchette1}.

For the sake of clarity, let us here consider that the measurement bases of $\mathcal{O}$ and $\Theta$, with $[\mathcal{O},\Theta] \neq 0$, belong to a set of finite dimension, i.e.\,that each measurement eigenvector $|\pi_{\mu_{\alpha}}\rangle$ admits only a finite number $d_{\pi}$ of elements $|\pi^{(j)}\rangle$. Then, the distribution of $l_{\theta_{m,j}}(\vec{o},\vec{\lambda})$ is obtained as following. By following a common procedure in LD theory, the terms of $l_{\theta_{m,j}}(\vec{o},\vec{\lambda})$ are grouped as a function of the number of times each dynamical transition probability $q_{j_{{\rm B}},j_{{\rm A}}}(\lambda) \equiv |\langle \pi^{(j_{{\rm B}})} |U(\lambda)| \pi^{(j_{{\rm A}})}\rangle|^2$ occurs, where the superscripts ${\rm B}$ and ${\rm A}$ stand, respectively, for ``Before'' and ``After'' the evolution via the propagator $U(\lambda)$. In this way, $l_{\theta_{m,j}}$ is recast in the following sum of i.i.d.\,random variables:
\begin{equation}
l_{\theta_{m,j}}(\vec{o},\vec{\lambda}) = \sum_{j_{{\rm B}},j_{{\rm A}}=1}^{d_{\pi}}\int_{\lambda}n_{j_{{\rm B}},j_{{\rm A}}}(\lambda)\ln q_{j_{{\rm B}},j_{{\rm A}}}(\lambda)d\lambda,
\end{equation}
with $n_{j_{{\rm B}},j_{{\rm A}}}(\lambda)$ denoting the relative frequencies for the occurrence of the $q_{j_{{\rm B}},j_{{\rm A}}}(\lambda)$'s,
usually different from the corresponding probability value $p_{j_{{\rm B}},j_{{\rm A}}}(\lambda)$. The deviation between them vanishes only by attaining a complete statistics for each admissible value of $q_{j_{{\rm B}},j_{{\rm A}}}(\lambda)$. This latter condition can be in principle realized, but likely a quite long system dynamical evolution could be required.

The probability distribution of a sum of $n$ i.i.d.\,random terms can be always written as an exponential, linearly decaying in $n$ with $n$ large. In particular, regarding $\textrm{Prob}(l_{\theta_{m,j}})$, it is given by an exponential distribution decaying in the number $m$ of projective measurements (see Appendix for more details), i.e.
\begin{equation}\label{Prob_l}
\textrm{Prob}(l_{\theta_{m,j}}(\vec{o},\vec{\lambda})) \asymp \exp(-m\,I(l_{\theta_{m,j}}(\vec{o},\vec{\lambda})/m)).
\end{equation}
In Eq.\,(\ref{Prob_l}), the function $I(l_{\theta_{m,j}}/m)$, also called the \emph{rate function} associated to the probability distribution $\textrm{Prob}(l_{\theta_{m,j}})$, equals to
\begin{equation}\label{rate_function}
I(l_{\theta_{m,j}}/m) \equiv \sum_{j_{{\rm B}},j_{{\rm A}}=1}^{d_{\pi}}\int_{\lambda}f_{j_{{\rm B}},j_{{\rm A}}}(\lambda)\ln\left(\frac{f_{j_{{\rm B}},j_{{\rm A}}}(\lambda)}
{p_{j_{{\rm B}},j_{{\rm A}}}(\lambda)}\right)d\lambda,
\end{equation}
where $f_{j_{{\rm B}},j_{{\rm A}}}(\lambda) \equiv n_{j_{{\rm B}},j_{{\rm A}}}(\lambda)/m$ for each set $(j_{{\rm B}},j_{{\rm A}},\lambda)$ of system parameters. The rate function $I(l_{\theta_{m,j}}/m)$ is the \emph{Kullback-Leibler distance} (or relative entropy) between the set $\{f_{j_{{\rm B}},j_{{\rm A}}}(\lambda)\}$ of \emph{scaled} relative frequencies and the set of probabilities $\{p_{j_{{\rm B}},j_{{\rm A}}}(\lambda)\}$, and, thus, has the properties to be positive and convex. Eq.\,(\ref{Prob_l}) is valid in the limit of $m$ large and indicates a \emph{unique nonequilibrium weighted partition} of the system configuration space (see the Appendix for the proof). For a small value of $m$, indeed, the distribution $\textrm{Prob}(l_{\theta_{m,j}})$ cannot be uniquely determined. The latter can be considered as a universal property of any dynamical evolutions given by the composition of quantum maps and projections. The alternative LD expression of Eq.\,(\ref{Prob_l}), providing the formal definition of $I(l_{\theta_{m,j}}/m)$, is
\begin{equation}\label{LD_definition}
\lim_{m\rightarrow\infty}-\frac{1}{m}\ln\,\textrm{Prob}(l_{\theta_{m,j}}(\vec{o},\vec{\lambda})) = I(l_{\theta_{m,j}}(\vec{o},\vec{\lambda})/m),
\end{equation}
where the number $m$ of measurements is assumed ideally infinite. If Eq.\,(\ref{LD_definition}), also called the \emph{large-deviation approximation}, holds, it means that the dominant behaviour of $\textrm{Prob}(l_{\theta_{m,j}})$ is \emph{convergent} and identically equal to a decaying exponential in $m$.

As last, through the contraction principle\,\cite{footnote4}, the distribution $\textrm{Prob}(p_{\theta_{m,j}}(\vec{o},\vec{\lambda}))$ is obtained:
\begin{small}
\begin{equation}
\textrm{Prob}(p_{\theta_{m,j}}(\vec{o},\vec{\lambda}))=\int \textrm{Prob}(l_{\theta_{m,j}}(\vec{o},\vec{\lambda}))
\delta(l_{\theta_{m,j}}-\ln p_{\theta_{m,j}})d\,l_{\theta_{m,j}}.
\end{equation}
\end{small}
Then, by applying the saddle point method\,\cite{BookWong1989}, one has that
\begin{equation}\label{Prob_p}
\textrm{Prob}(p_{\theta_{m,j}}(\vec{o},\vec{\lambda})) \asymp \exp(-m\,J(p_{\theta_{m,j}}(\vec{o},\vec{\lambda})/m)),
\end{equation}
where
\begin{equation}
J(p_{\theta_{m,j}}(\vec{o},\vec{\lambda})/m)\equiv\min_{l_{\theta_{j}}:\,l_{\theta_{j}}=\,\ln p_{\theta_{j}}} I(l_{\theta_{m,j}}(\vec{o},\vec{\lambda})/m).
\end{equation}
The result is that a quantum system exhibiting stochastic evolutions (e.g.\,due to random interactions with the environment) and repeatedly monitored by an observer tends to reach in probability a unique configuration defined by specific probability distributions of its characteristic parameters.

\subsection{Most probable distribution}

The expression of the most probable distribution of $\Theta$'s outcomes is here discussed. In general, from the knowledge of the rate function $I(\xi/m)$ associated to $\textrm{Prob}(\xi)$, we can then compute the most probable value $\xi^{\star}$, representing the best prediction for the random variable $\xi$ in a single realization of the system dynamics. Specifically, the most probable value of the log-conditional-probability $l_{\theta_{m,j}}(\vec{o},\vec{\lambda})$, i.e., $l^{\star}_{\theta_{m,j}}$, is obtained by evaluating the value at which the rate function $I(l_{\theta_{m,j}}/m)$ is minimized as a function of $l_{\theta_{m,j}}$. As proved in the Appendix, the closed-form expression of $l^{\star}_{\theta_{m,j}}$ is
\begin{equation}\label{l_star}
l^{\star}_{\theta_{m,j}} = m\sum_{j_{{\rm B}},j_{{\rm A}}=1}^{d_{\pi}}\int_{\lambda}p_{j_{{\rm B}},j_{{\rm A}}}(\lambda)\ln q_{j_{{\rm B}},j_{{\rm A}}}(\lambda)\,d\lambda,
\end{equation}
corresponding to the statement that $n_{j_{{\rm B}},j_{{\rm A}}}(\lambda) = m\,p_{j_{{\rm B}},j_{{\rm A}}}(\lambda)$ for every $(j_{{\rm B}},j_{{\rm A}},\lambda)$. This means that for large $m$ the most probable trajectories of the system dynamics are those allowing for the equality between the scaled relative frequencies $f_{j_{{\rm B}},j_{{\rm A}}}(\lambda)$ and the occurrence probabilities $p_{j_{{\rm B}},j_{{\rm A}}}(\lambda)$. Only this condition minimizes the relative Shannon entropy between the configurations induced by the stochastic dynamics of the system. Once more, it is worth observing the importance of imposing $m\rightarrow\infty$; indeed, the value of the scaled relative frequencies $f_{j_{{\rm B}},j_{{\rm A}}}$ can get closer to that of the probabilities $p_{j_{{\rm B}},j_{{\rm A}}}$ only if the number of configurations generated by the stochasticity within the dynamics is as large as possible. Finally, to derive the most probable value of the conditional probabilities $p^{\star}_{\theta_{m,j}}$, we use again the contraction principle from LD theory, but this time on the functional relation between $l_{\theta_{m,j}}$ and $p_{\theta_{m,j}}$, i.e.\,$p_{\theta_{m,j}} = e^{l_{\theta_{m,j}}}$:
\begin{equation}\label{eq_p_star}
p^{\star}_{\theta_{m,j}} = \exp\left(m\sum_{j_{{\rm B}},j_{{\rm A}}=1}^{d_{\pi}}\int_{\lambda}p_{j_{{\rm B}},j_{{\rm A}}}(\lambda)\ln q_{j_{{\rm B}},j_{{\rm A}}}(\lambda)\,d\lambda\right).
\end{equation}

\section{Discussion}

Having proved that the nonequilibrium statistics of an arbitrary observable obeys the LD principle for a sufficiently large number of intermediate projection events, the answer to the original question is given by the distribution of $\Theta$'s outcomes with occurrence conditional probabilities equal to $p^{\star}_{\theta_{m,j}}$. Such a distribution is denoted as the \emph{most probable distribution} (see e.g.\,Fig.\,\ref{fig:fig2} inset (a)):
\begin{equation}\label{Prob_star_theta}
\textrm{Prob}^{\star}(\theta) \equiv \frac{1}{\mathcal{N}}\sum_{j}\delta(\theta - \theta_{m,j})p^{\star}_{\theta_{m,j}},
\end{equation}
which has to be normalized by the factor $\mathcal{N}$ so as to ensure that $\int \textrm{Prob}^{\star}(\theta)d\theta = 1$. Accordingly, the predictions about the result of single-shot measurements from $\Theta$ according to the most probable distribution $\textrm{Prob}^{\star}(\theta)$ are expected to be more accurate than the ones that we would obtain by directly computing the expectation value of the measurement outcomes. This is because \emph{large fluctuations} within the evolution of the system are now properly weighted and thus correctly included in the outcome distributions. In this regard, also quantum noise sensing techniques\,\cite{DegenReview2017} could be used, so as to improve the a-priori information on $p(\lambda)$.

In conclusion, being the average $\overline{p}_{\theta_{m,j}}$ explicitly equal to
\begin{equation}
\langle p_{\theta_{m,j}}(\vec{o},\vec{\lambda})\rangle
= \left(\sum_{j_{{\rm B}},j_{{\rm A}}=1}^{d_{\pi}}\int_{\lambda}p_{j_{{\rm B}},j_{{\rm A}}}(\lambda)q_{j_{{\rm B}},j_{{\rm A}}}(\lambda)d\lambda\right)^{m},
\end{equation}
one can observe a deviation between the most probable values $p^{\star}_{\theta_{m,j}}$ and the probabilities $\overline{p}_{\theta_{m,j}}$ to measure on average the outcome $\theta_{m,j}$ at the final time instant $t_{m}$. Specifically, by applying Jensen's inequality\,\cite{footnote2} to $p^{\star}_{\theta_{m,j}}$ and $\overline{p}_{\theta_{m,j}}$,
the inequality $p^{\star}_{\theta_{m,j}} \leq \overline{p}_{\theta_{m,j}}$ is obtained. The latter well justifies the introduction of the normalization factor $\mathcal{N}$ in Eq.\,(\ref{Prob_star_theta}). In fact, the normalization of $\textrm{Prob}^{\star}(\theta)$ has to be ensured, and the values of the $p^{\star}_{\theta_{m,j}}$'s are thus corrected by dividing for $\mathcal{N} \equiv \sum_{j}p^{\star}_{\theta_{m,j}} \leq 1$, so that $p^{\star}_{\theta_{m,j}}/\mathcal{N} = 1 - (\sum_{k,\,k\neq j}p^{\star}_{\theta_{m,k}})/\mathcal{N}$, $\forall\,j=1,\ldots,\dim(\mathcal{H})$.

\section{Experimental proposal}

\begin{figure*}
  \includegraphics[width=0.9\textwidth,height=0.4\textwidth]{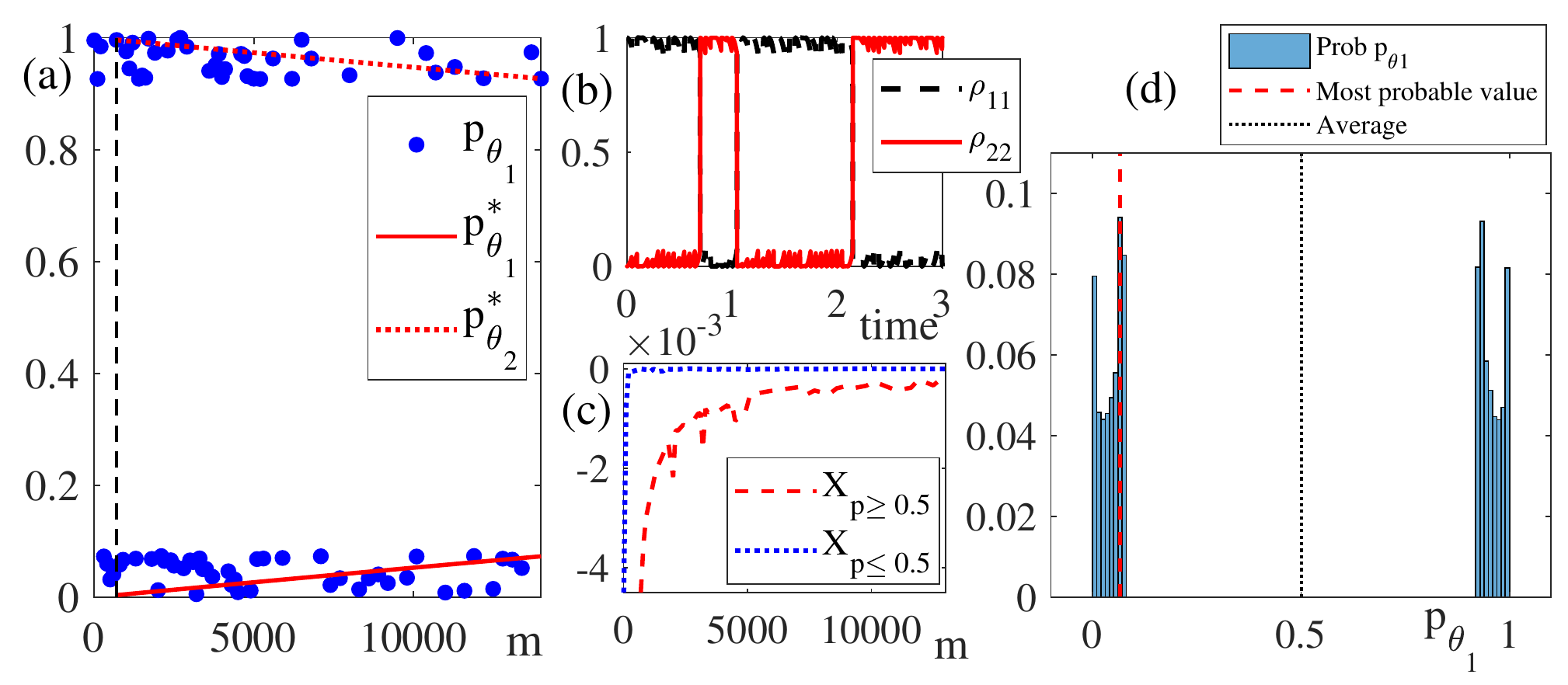}
  \caption{Experimental proposal: Exact quantum observable statistics of a 2LS with stochastic Hamiltonian. (a) Theoretical most probable values $p_{\theta_1}^{\star}$ and $p_{\theta_2}^{\star}$ as a function of $m$ compared with different values of the probability $p_{\theta_1}$, which has been numerically obtained from single realizations of the system transformation. (b) Time-behaviour of system populations $\rho^{(11)}$ and $\rho^{(22)}$ by numerically solving the stochastic evolution of the 2LS for one realization of $\vec{o}$ and $\vec{\varphi}$. (c) Plot of $\ln(p_{\theta_1})/m$ from data of inset (a). (d) Histogram of $p_{\theta_1}$, with $m=14\cdot10^{3}$, over $20\cdot10^{3}$ repetitions of the system dynamics.}
  \label{fig:fig2}
\end{figure*}

Let us consider a two-level system (2LS) with Hamiltonian $H=\frac{1}{2}\omega(\cos\varphi\sigma_{z}-\sin\varphi\sigma_{x})$, with $\omega=11$ (in natural units s.t.\,$\hbar=1$) and $\sigma_{x},\sigma_{z}$ Pauli matrices. This system, realized e.g.\,within a Bose-Einstein condensate of Rubidium atoms confined by a magnetic micro-trap as in\,\cite{GherardiniErgodicity}, is intrinsically stochastic due to random fluctuations of the angle $\varphi$. They can be produced for example by a random coupling between the system and the laser source that coherently drives it. In particular, we assume sampling of $\varphi$ from the Rayleigh probability distribution $p(\varphi)=(\varphi/\sigma^2)\exp(-\varphi^{2}/(2\sigma^2))$ with scale parameter $\sigma=1$. Then, the 2LS, starting from the ground $|0\rangle$, is monitored by a sequence of $m$ repeated projective measurements. They are separated by the constant time interval $\tau=0.05$ (in natural units) and defined by the projectors $\Pi_1 \equiv |0\rangle\langle0|$ and $\Pi_2 \equiv |1\rangle\langle1|$. After each dynamical evolution with whole duration $t_{m}=m\tau$, the quantum observable
\begin{equation}
\Theta \equiv \sum_{j=1}^{2}\theta_j\Pi_{\theta_j} = \rho^{(11)}\Pi_1 + \rho^{(22)}\Pi_2
\end{equation}
is measured, where the measurement outcomes $\theta_1 \equiv \rho^{(11)}$ and $\theta_2 \equiv \rho^{(22)}=1-\rho^{(11)}$ identify the percentage of system population respectively in the ground and excited states, $|0\rangle$ and $|1\rangle$ respectively. Notice that, being $\mathcal{S}$ a 2LS, we analyze the fluctuations of only the conditional probability $p_{\theta_1}(\vec{o},\vec{\varphi})\equiv{\rm Tr}[\Pi_{\theta_1}\rho_{m,\vec{o},\vec{\varphi}}]$.

In Fig.\,\ref{fig:fig2}-(a) we show the most probable values of $p_{\theta_1}$, i.e.\,$p_{\theta_1}^{\star}$ (solid red curve), which has been computed by plotting the analytical expression of Eq.\,(\ref{eq_p_star}) as a function of $m$. The following procedure has been followed. First, we have numerically computed the probabilities $p_{\theta_1}$ (blue points in Fig.\,\ref{fig:fig2}-(a)) to measure the measurement outcome $\theta_{1}$ after each realization of the system stochastic transformation (see inset (b)). Then, we have derived the quantity $\ln(p_{\theta_1})/m$, by distinguishing between values of $p_{\theta_1}$ smaller (dotted blue curve -- inset (c)) and greater (dashed red curve -- inset (c)) than $1/2$. As it can be observed, only the dotted blue curve is practically constant for $m \geq 700$ and equal to $X \simeq -4.22\cdot 10^{-6}$. In particular, from the theoretical findings, we can state that for $m$ large $\ln(p_{\theta_1}^{\star}) \asymp mX$ (solid red curve -- inset (a)) with
\begin{eqnarray}
&\displaystyle{X \equiv \sum_{j_{\rm B},j_{\rm A}}\int p_{j_{\rm B},j_{\rm A}}(\varphi)\ln(q_{j_{\rm B},j_{\rm A}}(\varphi))d\varphi},\,\,\,\text{i.e.,}& \\
&\displaystyle{X = \frac{1}{N_{\rm meas}}\int d\varphi p(\varphi)\sum_{k,j=0}^{1}\ln|\langle k|U(\varphi)|j\rangle|^{2}}&
\end{eqnarray}
for this specific example, where $U(\varphi) \equiv \exp(-iH(\varphi)\tau)$ and $N_{\rm meas}$ encodes the occurrence probabilities of the intermediate quantum measurement projectors. As a further test, we have numerically obtained the probability distribution of $p_{\theta_1}$ (inset (d)) with $m=14\cdot10^{3}$, by repeating $20\cdot10^{3}$ times the dynamics of the system, and computed the value of $p_{\theta_1}^{\rm peak}$ at which ${\rm Prob}(p_{\theta_1})$ admits the peak; we measured $p_{\theta_1}^{\rm peak}=0.0736$ and an average value $\overline{p}_{\theta_{1}}=0.5016$. This value should equal to the most probable value $p_{\theta_1}^{\star}$ and this has been effectively verified by the numerics, since from the solid red curve in Fig.\,\ref{fig:fig2}-(a) we find for $m=14\cdot10^{3}$ the value $p_{\theta_1}^{\star}=0.0731$.

\section{Conclusions}

These results combine out-of-equilibrium quantum systems with statistical mechanics methods: They are expected to be a tool to defeat noise in a quantum system, since they provide predictions of the stochastic evolution of the system after the single realization.

The author gratefully acknowledges F.\,Caruso and M.\,Artoni for a detailed reading of the paper, A.\,Trombettoni and M.\,Fattori for several useful comments, and Shamik Gupta for insightful discussions on LD theory, in particular on the derivation of the measurement outcomes'probability distribution. S.G.\,was financially supported from the Fondazione CR Firenze through the project Q-BIOSCAN.

\section*{Appendix}

\subsection*{I.\,Measurement conditional probabilities}

Let us express the measurement projectors $\Pi_{\mu_{\alpha}}$ applied at time instants $t_{\alpha}$, $\alpha = 1,\ldots,m$, as a function of their eigenvectors $|\pi_{\mu_{\alpha}}\rangle$, i.e.\,$\Pi_{\mu_{\alpha}} \equiv |\pi_{\mu_{\alpha}}\rangle\langle\pi_{\mu_{\alpha}}|$, with $\mu \in \{\theta,o\}$. Thus, by substituting the definition of the measurement super-operators $\mathcal{P}_{\theta_{m}}$ and $\mathcal{P}_{o_{\alpha}}$ into the equation
\begin{equation}
p_{\theta_{m}}(\vec{o},\vec{\lambda}) \equiv \textrm{Tr}\left[
\mathcal{P}_{\theta_{m}}\mathcal{U}_{m}\mathcal{P}_{o_{m-1}}\mathcal{U}_{m-1}\cdots\mathcal{P}_{o_1}\mathcal{U}_{1}[\rho_0]\right],
\end{equation}
the conditional probability $p_{\theta_{m,j}}(\vec{o},\vec{\lambda})$ can be written as the product of the transition probabilities (random terms) that the quantum state moves from $|\pi_{\mu_{\alpha-1}}\rangle$ to $|\pi_{\mu_{\alpha}}\rangle$ via the unitary operator $U_{\alpha}(\lambda_{\alpha})$:
\begin{widetext}
\begin{eqnarray}
&& p_{\theta_{m,j}}(\vec{o},\vec{\lambda})
= \textrm{Tr}[\Pi_{\theta_{m,j}}U_{m}\Pi_{o_m-1}U_{m-1}\cdots\Pi_{o_1}U_1\rho_{0}U_{1}^{\dagger}\Pi_{o_1}\cdots U_{m-1}^{\dagger}\Pi_{o_{m-1}}U_{m}^{\dagger}\Pi_{\theta_{m,j}}]\nonumber \\
&& = \textrm{Tr}\left[|\pi_{\theta_{m,j}}\rangle\langle\pi_{\theta_{m,j}}|U_{m}|\pi_{o_{m-1}}\rangle\langle\pi_{o_{m-1}}|U_{m-1}\cdots|\pi_{o_1}\rangle\langle\pi_{o_1}|U_{1}
\rho_{0}U_{1}^{\dagger}|\pi_{o_1}\rangle\langle\pi_{o_1}|\cdots U_{m-1}^{\dagger}|\pi_{o_{m-1}}\rangle\langle\pi_{o_{m-1}}|U_{m}^{\dagger}|\pi_{\theta_{m,j}}\rangle\langle\pi_{\theta_{m,j}}|\right]\nonumber \\
&& = \langle\pi_{o_1}|U_1\rho_{0}U_{1}^{\dagger}|\pi_{o_1}\rangle \cdot \prod_{k=2}^{m-1}|\langle\pi_{o_{k-1}}|U_k|\pi_{o_k}\rangle|^2 \cdot |\langle\pi_{o_{m-1}}|U_{m}|\pi_{\theta_{m,j}}\rangle|^2.
\end{eqnarray}
\end{widetext}
$\rho_0$ is the initial density matrix of the open quantum system $\mathcal{S}$ before the transformation induced by random interactions with the environment $\mathcal{E}$ and the monitoring by an observer. Moreover, we also assume that $\rho_0$ is defined by the pure state $|\pi_{0}\rangle$, i.e., $\rho_0 \equiv |\pi_{0}\rangle\langle\pi_{0}|$. For example this assumption is verified when the measurement outcomes at the final time $t_{m}$ are obtained by a two-time measurement scheme \cite{CampisiPRL2010}. In such a case, indeed, $\rho_0$ is the density matrix of $\mathcal{S}$ after the first measurement of the scheme, and thus it is described only by a pure state. Therefore, under this further hypothesis, one has that $\langle\pi_{o_1}|U_1\rho_{0}U_{1}^{\dagger}|\pi_{0_1}\rangle = |\langle\pi_0|U_1|\pi_{0_1}\rangle|^2$, so that
\begin{equation}
p_{\theta_{m,j}}(\vec{o},\vec{\lambda}) = \prod_{\alpha=1}^{m}|\langle\pi_{\mu_{\alpha-1}}|U_{\alpha}(\lambda_{\alpha})|\pi_{\mu_{\alpha}}\rangle|^{2},
\end{equation}
with $\pi_{\mu_0} \equiv \pi_0$, $\mu_{\alpha} \equiv o_{\alpha}$ for $\alpha = 1,\ldots,m-1$ and $\pi_{\mu_{m}} \equiv \pi_{\theta_{m,j}}$.

\subsection*{II.\,Derivation of $\textrm{Prob}(l_{\theta_{m,j}}(\vec{o},\vec{\lambda}))$ via LD theory}

The log-probability $l_{\theta_{m,j}}(\vec{o},\vec{\lambda})$ is the logarithm of the conditional probability $p_{\theta_{m,j}}(\vec{o},\vec{\lambda})$ defining the probability to obtain the $j$-th measurement outcome $\theta_{m,j}$ at the time instant $t_{m}$, conditioned to the specific realizations of the sequences $\vec{o}$ and $\vec{\lambda}$. More formally,
\begin{eqnarray}
l_{\theta_{m,j}}(\vec{o},\vec{\lambda}) &\equiv& \ln p_{\theta_{m,j}}(\vec{o},\vec{\lambda}) = \sum_{\alpha = 1}^{m} \ln q_{\alpha}(\mu_{\alpha-1},\mu_{\alpha},\lambda_{\alpha})\nonumber \\
&=& \sum_{\alpha = 1}^{m} \ln \left(|\langle\pi_{\mu_{\alpha-1}}|U_{\alpha}(\lambda_{\alpha})|\pi_{\mu_{\alpha}}\rangle|^2\right).
\end{eqnarray}
For the sake of clarity, let us assume the following two hypotheses: (i) $p(\lambda)$ is assumed to be a $d_{\lambda}-$dimensional Bernoulli distribution, with the result that at each time instant $t_{\alpha}$ the parameter $\lambda_{\alpha}$ takes on $d_{\lambda}$ possible values $\lambda^{(1)},\ldots,\lambda^{(d_{\lambda})}$ with corresponding probabilities $p_{\lambda}^{(1)},\ldots,p_{\lambda}^{(d_{\lambda})}$ so that $\sum_{i = 1}^{d_{\lambda}}p_{\lambda}^{(i)} = 1$. The index $i$ denotes the values that can be assumed by $\lambda$. (ii) The measurement bases of $\Theta$ and $\mathcal{O}$, given by the set of eigenvectors $\{|\pi_{\mu_{\alpha}}\rangle\}$, belong to a set of finite dimension and this means that each ket $|\pi_{\mu_{\alpha}}\rangle$ admits only a finite number $d_{\pi}$ of elements $|\pi^{(j)}\rangle$ with probability $p_{\pi}$.
Notice that the index for the dimensionality of the measurement basis configurations is $j$, as well as that for the dimension of $\mathcal{H}$. By following a common procedure in LD theory, we group the terms of $l_{\theta_{m,j}}(\vec{o},\vec{\lambda})$ as a function of the number of times (relative frequencies) each dynamical transition probability
$q_{j_{{\rm B}},i,j_{{\rm A}}} \equiv |\langle \pi^{(j_{{\rm B}})}|U(\lambda^{(i)})|\pi^{(j_{{\rm A}})}\rangle|^{2}$
occurs during the transformation of the system.
We denote with $n_{j_{{\rm B}},i,j_{{\rm A}}}$ the relative frequencies for the occurrence of the $q_{j_{{\rm B}},i,j_{{\rm A}}}$'s. The corresponding occurrence probabilities, instead, are denoted as $p_{j_{{\rm B}},i,j_{{\rm A}}}$. Over many realizations of the system transformation, one can reasonably assume that $p_{\pi}$ is sampled by a uniform distribution, such that $p_{\pi} = 1/d_{\pi}$ and
$p_{j_{{\rm B}},i,j_{{\rm A}}} \equiv p_{\pi}^{(j_{{\rm B}})}p_{\lambda}^{(i)}p_{\pi}^{(j_{{\rm A}})} = p_{\lambda}^{(i)}/d_{\pi}^{2}$. The latter procedure directly leads to the analytical expression of $l_{\theta_{m,j}}(\vec{o},\vec{\lambda})$, i.e.,
\begin{equation}\label{logP_SI}
l_{\theta_{m,j}}(\vec{o},\vec{\lambda}) = \sum_{j_{{\rm B}}=1}^{d_{\pi}}\sum_{i=1}^{d_{\lambda}}\sum_{j_{{\rm A}}=1}^{d_{\pi}}n_{j_{{\rm B}},i,j_{{\rm A}}}\ln q_{j_{{\rm B}},i,j_{{\rm A}}}.
\end{equation}
Eq.\,(\ref{logP_SI}) shows us that $l_{\theta_{m,j}}(\vec{o},\vec{\lambda})$ can be written as the sum of the i.i.d.\,dynamical transition probabilities $q_{j_{{\rm B}},i,j_{{\rm A}}}$, weighted by the relative frequencies $n_{j_{{\rm B}},i,j_{{\rm A}}}$ corresponding to the occurrence statistics of the $q_{j_{{\rm B}},i,j_{{\rm A}}}$'s. Notice that the relative frequencies $n_{j_{{\rm B}},i,j_{{\rm A}}}$ are usually different from the corresponding probability values $p_{j_{{\rm B}},i,j_{{\rm A}}}$. Then, by taking Eq.\,(\ref{logP_SI}), the distribution probability of $l_{\theta_{m,j}}(\vec{o},\vec{\lambda})$ is given by
\begin{eqnarray}\label{prob_l_SM}
&\textrm{Prob}(l_{\theta_{m,j}}(\vec{o},\vec{\lambda})) = \frac{m!}{\displaystyle{\prod_{j_{{\rm B}},i,j_{{\rm A}}}n_{j_{{\rm B}},i,j_{{\rm A}}}!}}\prod_{j_{{\rm B}},i,j_{{\rm A}}}(p_{j_{{\rm B}},i,j_{{\rm A}}})^{n_{j_{{\rm B}},i,j_{{\rm A}}}}&\nonumber \\
&\times\delta\left(l_{\theta_{m,j}}(\vec{o},\vec{\lambda}) - \sum_{j_{{\rm B}},i,j_{{\rm A}}}n_{j_{{\rm B}},i,j_{{\rm A}}}\ln q_{j_{{\rm B}},i,j_{{\rm A}}}\right),&
\end{eqnarray}
where $\delta(\cdot)$ denotes the Kronecker delta. Let us observe that in Eq.\,(\ref{prob_l_SM}), to simplify the notation, we have used the symbols $\displaystyle{\sum_{j_{{\rm B}},i,j_{{\rm A}}}}$ and $\displaystyle{\prod_{j_{{\rm B}},i,j_{{\rm A}}}}$ to denote respectively $\displaystyle{\sum_{j_{{\rm B}}=1}^{d_{\pi}}\sum_{i=1}^{d_{\lambda}}\sum_{j_{{\rm A}}=1}^{d_{\pi}}}$ and $\displaystyle{\prod_{j_{{\rm B}}=1}^{d_{\pi}}\prod_{i=1}^{d_{\lambda}}\prod_{j_{{\rm A}}=1}^{d_{\pi}}}$. Then, by imposing in Eq.\,(\ref{prob_l_SM}) the condition given by the Kronecker delta, one has that
\begin{equation}\label{prob_distr_l}
\textrm{Prob}(l_{\theta_{m,j}}(\vec{o},\vec{\lambda})) = \frac{m!}{\displaystyle{\prod_{j_{{\rm B}},i,j_{{\rm A}}}\widehat{n}_{j_{{\rm B}},i,j_{{\rm A}}}!}}\prod_{j_{{\rm B}},i,j_{{\rm A}}}(p_{j_{{\rm B}},i,j_{{\rm A}}})^{\widehat{n}_{j_{{\rm B}},i,j_{{\rm A}}}},
\end{equation}
where the relative frequencies $\widehat{n}_{j_{{\rm B}},i,j_{{\rm A}}}$'s have to satisfy the following constraints:
\begin{equation}\label{constraints_1_e_2}
\begin{cases}
\displaystyle{\sum_{j_{{\rm B}},i,j_{{\rm A}}}n_{j_{{\rm B}},i,j_{{\rm A}}} = m} \\
\displaystyle{l_{\theta_{m,j}}(\vec{o},\vec{\lambda}) = \sum_{j_{{\rm B}},i,j_{{\rm A}}}n_{j_{{\rm B}},i,j_{{\rm A}}}\ln q_{j_{{\rm B}},i,j_{{\rm A}}}}.
\end{cases}
\end{equation}
By combining together the constraints (\ref{constraints_1_e_2}), we obtain a unique constraint equation for $l_{\theta_{m,j}}$, i.e.,
\begin{equation}\label{unique_constraint}
l_{\theta_{m,j}}(\vec{o},\vec{\lambda}) = m\ln q_{d_{\pi},d_{\lambda},d_{\pi}} -
\widetilde{\sum_{j_{{\rm B}},i,j_{{\rm A}}}}\widehat{n}_{j_{{\rm B}},i,j_{{\rm A}}}\gamma_{j_{{\rm B}},i,j_{{\rm A}}},
\end{equation}
where
\begin{equation}\label{app_gamma}
\gamma_{j_{{\rm B}},i,j_{{\rm A}}} \equiv \frac{\ln q_{d_{\pi},d_{\lambda},d_{\pi}}}{\ln q_{j_{{\rm B}},i,j_{{\rm A}}}}.
\end{equation}
It is worth observing that in deriving the constraint (\ref{unique_constraint}) we have chosen $(d_{\pi},d_{\lambda},d_{\pi})$ as the \emph{reference triplet} of the configuration space that defines the stochastic trajectory of the system dynamics in a single realization. This choice is arbitrary and represents a degree of freedom of the procedure. However, that is not surprising because the number of constraints of formula\,(\ref{constraints_1_e_2}) is smaller than the number of relative frequencies $n_{j_{{\rm B}},i,j_{{\rm A}}}$, so that the values of $\widehat{n}_{j_{{\rm B}},i,j_{{\rm A}}}$ that satisfy Eq.\,(\ref{prob_distr_l}) are generally not uniquely determined. This means that in order to obtain a unique analytical expression of $\textrm{Prob}(l_{\theta_{m,j}}(\vec{o},\vec{\lambda}))$, we need to answer the following questions: Which are the \emph{unique} values of the relative frequencies $\widehat{n}_{j_{{\rm B}},i,j_{{\rm A}}}$ obeying the constraint equation (\ref{unique_constraint})? By generalizing the results in \cite{GherardiniNJP}, we can prove that \emph{there exists a unique value for the $\widehat{n}_{j_{{\rm B}},i,j_{{\rm A}}}$'s, under the hypothesis of a sufficiently large number $m$ of intermediate projective measurements}. To see this, let us consider the product $\widehat{n}_{a,b,c}\gamma_{a,b,c}$ with generic indices $(a,b,c)$:
\begin{eqnarray}
&\displaystyle{\widehat{n}_{a,b,c}\gamma_{a,b,c} = m\left(\ln q_{d_{\pi},d_{\lambda},d_{\pi}} - l_{\theta_{m,j}}(\vec{o},\vec{\lambda})/m\right)}& \nonumber \\
&-\displaystyle{\widetilde{\sum_{j_{{\rm B}},i,j_{{\rm A}};\,\,(j_{{\rm B}},i,j_{{\rm A}})\neq (a,b,c)}}\widehat{n}_{j_{{\rm B}},i,j_{{\rm A}}}\gamma_{j_{{\rm B}},i,j_{{\rm A}}}},&
\end{eqnarray}
i.e.,
\begin{eqnarray}
&\displaystyle{\frac{\widehat{n}_{a,b,c}}{m} = \frac{\ln q_{d_{\pi},d_{\lambda},d_{\pi}} - l_{\theta_{m,j}}(\vec{o},\vec{\lambda})/m}{\gamma_{a,b,c}}}& \nonumber \\
&-\displaystyle{\widetilde{\sum_{j_{{\rm B}},i,j_{{\rm A}};\,\,j_{{\rm B}}\neq a,i\neq b,j_{{\rm A}}\neq c}}\widehat{n}_{j_{{\rm B}},i,j_{{\rm A}}}\frac{\widehat{n}_{j_{{\rm B}},i,j_{{\rm A}}}}{m}\frac{\gamma_{j_{{\rm B}},i,j_{{\rm A}}}}{\gamma_{a,b,c}}}&.
\end{eqnarray}
Being $\widehat{n}_{j_{{\rm B}},i,j_{{\rm A}}}$'s relative frequencies, it still holds that
$\lim_{m \rightarrow \infty}\frac{\widehat{n}_{j_{{\rm B}},i,j_{{\rm A}}}}{m} = 0$,
for each triplet $(j_{{\rm B}},i,j_{{\rm A}})$ except $(a,b,c)$, with the result that
\begin{eqnarray}
&\displaystyle{\frac{\widehat{n}_{a,b,c}}{m} \asymp \frac{\ln q_{d_{\pi},d_{\lambda},d_{\pi}} - l_{\theta_{m,j}}(\vec{o},\vec{\lambda})/m}{\gamma_{a,b,c}}}& \nonumber \\
&=\displaystyle{\frac{1}{N_{\rm sf}}\left(\frac{\ln q_{d_{\pi},d_{\lambda},d_{\pi}} - l_{\theta_{m,j}}(\vec{o},\vec{\lambda})/m}{\gamma_{a,b,c}}\right)}&,
\end{eqnarray}
with $N_{\rm sf}$ denoting the corresponding \emph{scaling factor}. Thus, this means that for \emph{large} $m$
\begin{equation}\label{hat_n_times_gamma}
\widehat{n}_{a,b,c}\gamma_{a,b,c} \asymp \frac{m\ln q_{d_{\pi},d_{\lambda},d_{\pi}} - l_{\theta_{m,j}}(\vec{o},\vec{\lambda})}{N_{\rm sf}} = {\rm constant},
\end{equation}
for each triplet $(a,b,c)$. Eq.\,(\ref{hat_n_times_gamma}) is a quite important result, since it denotes the existence of a \emph{unique nonequilibrium weighted partition} of the system configuration space, once we have fixed the reference triplet $(d_{\pi},d_{\lambda},d_{\pi})$. Such a property is thus the key point for the derivation of $\textrm{Prob}(l_{\theta_{m,j}}(\vec{o},\vec{\lambda}))$. In particular, by summing together the terms $\widehat{n}_{a,b,c}\gamma_{a,b,c}$ over all the possible values that can be assumed by $(a,b,c)$ except for the triplet $(d_{\pi},d_{\lambda},d_{\pi})$, we find that
\begin{small}
\begin{equation}\label{equipartition_SM}
\widetilde{\sum_{j_{{\rm B}},i,j_{{\rm A}}}}\widehat{n}_{j_{{\rm B}},i,j_{{\rm A}}}\gamma_{j_{{\rm B}},i,j_{{\rm A}}}
\asymp \frac{(d_{\rm tot} - 1)}{N_{\rm sf}}\left(m\ln q_{d_{\pi},d_{\lambda},d_{\pi}} - l_{\theta_{m,j}}(\vec{o},\vec{\lambda})\right),
\end{equation}
\end{small}
where $d_{\rm tot} \equiv 2d_{\pi} + d_{\lambda}$ is the dimension of the statistical ensemble defining the stochastic transformation of the system from $t_0$ to $t_{m}$. Therefore, by comparing Eqs.\,(\ref{unique_constraint}) and (\ref{equipartition_SM}), one can state that $N_{\rm sf} = d_{\rm tot} - 1$, so that for $m$ sufficiently large
\begin{equation}\label{n_tilde}
\widehat{n}_{a,b,c} \asymp \frac{m\ln q_{d_{\pi},d_{\lambda},d_{\pi}} - l_{\theta_{m,j}}(\vec{o},\vec{\lambda})}{(d_{\rm tot} - 1)\gamma_{a,b,c}}.
\end{equation}
Once we have obtained a closed solution for the value of the relative frequencies obeying the constraints (\ref{constraints_1_e_2}), we are able to validate the exponential approximation given by the \emph{large-deviation principle} for the probability distribution of $l_{\theta_{m,j}}$ in the thermodynamic limit of $m\rightarrow\infty$.
To practically derive the LD form of $\textrm{Prob}(l_{\theta_{m,j}}(\vec{o},\vec{\lambda}))$, just take Eq.\,(\ref{prob_distr_l}) and apply the Stirling approximation on $\ln(m!)$ and $\ln(\widehat{n}_{j_{{\rm B}},i,j_{{\rm A}}}!)$, which is valid again in the limit of large $m$:
\begin{widetext}
\begin{eqnarray}
\textrm{Prob}(l_{\theta_{m,j}}(\vec{o},\vec{\lambda})) &=& \exp\left(\ln(m!) - \sum_{j_{{\rm B}},i,j_{{\rm A}}}\ln(\widehat{n}_{j_{{\rm B}},i,j_{{\rm A}}}!)
+ \sum_{j_{{\rm B}},i,j_{{\rm A}}}\widehat{n}_{j_{{\rm B}},i,j_{{\rm A}}}\ln p_{j_{{\rm B}},i,j_{{\rm A}}}\right)\nonumber \\
&\asymp& \exp\left(m\ln m - m - \sum_{j_{{\rm B}},i,j_{{\rm A}}}\widehat{n}_{j_{{\rm B}},i,j_{{\rm A}}}\ln \widehat{n}_{j_{{\rm B}},i,j_{{\rm A}}}+\sum_{j_{{\rm B}},i,j_{{\rm A}}}\widehat{n}_{j_{{\rm B}},i,j_{{\rm A}}}+
\sum_{j_{{\rm B}},i,j_{{\rm A}}}\widehat{n}_{j_{{\rm B}},i,j_{{\rm A}}}\ln p_{j_{{\rm B}},i,j_{{\rm A}}}\right)\nonumber \\
&=& \exp\left(m\ln m-\sum_{j_{{\rm B}},i,j_{{\rm A}}}\widehat{n}_{j_{{\rm B}},i,j_{{\rm A}}}\ln \widehat{n}_{j_{{\rm B}},i,j_{{\rm A}}}+\sum_{j_{{\rm B}},i,j_{{\rm A}}}\widehat{n}_{j_{{\rm B}},i,j_{{\rm A}}}\ln p_{j_{{\rm B}},i,j_{{\rm A}}}\right).
\end{eqnarray}
\end{widetext}
Then, by substituting the expression of $\widehat{n}_{j_{{\rm B}},i,j_{{\rm A}}}$'s given by Eq.\,(\ref{n_tilde}), after straightforward calculations one finds that
\begin{equation}
\textrm{Prob}(l_{\theta_{m,j}}(\vec{o},\vec{\lambda})) \asymp \exp(-m\,I(l_{\theta_{m,j}}(\vec{o},\vec{\lambda})/m)),
\end{equation}
where
\begin{equation}\label{app_rate_function}
I(l_{\theta_{m,j}}(\vec{o},\vec{\lambda})/m) = \sum_{j_{{\rm B}}=1}^{d_{\pi}}\sum_{i=1}^{d_{\lambda}}\sum_{j_{{\rm A}}=1}^{d_{\pi}}f_{j_{{\rm B}},i,j_{{\rm A}}}\ln\left(\frac{f_{j_{{\rm B}},i,j_{{\rm A}}}}
{p_{j_{{\rm B}},i,j_{{\rm A}}}}\right)
\end{equation}
is the rate function associated to the probability distribution $\textrm{Prob}(l_{\theta_{m,j}}(\vec{o},\vec{\lambda}))$. In particular, in Eq.\,(\ref{app_rate_function}),
\begin{equation}\label{app_f}
f_{j_{{\rm B}},i,j_{{\rm A}}} \equiv \frac{\ln q_{d_{\pi},d_{\lambda},d_{\pi}} - l_{\theta_{m,j}}(\vec{o},\vec{\lambda})/m}{(d_{\rm tot}-1)\gamma_{j_{{\rm B}},i,j_{{\rm A}}}}
\end{equation}
for each triplet $(j_{{\rm B}},i,j_{{\rm A}}) \neq (d_{\pi},d_{\lambda},d_{\pi})$, while
\begin{equation}
f_{d_{\pi},d_{\lambda},d_{\pi}} \equiv 1 - \widetilde{\sum_{j_{{\rm B}},i,j_{{\rm A}}}}f_{j_{{\rm B}},i,j_{{\rm A}}}.
\end{equation}
As a final remark, given the generic triplet $(a,b,c)$, it is worth noting that $f_{a,b,c}$ and the relative frequency $\widehat{n}_{a,b,c}$ are simply related by the following equation:
\begin{equation}
f_{a,b,c} = \frac{\widehat{n}_{a,b,c}}{m}\,.
\end{equation}

\subsection*{III.\,Most probable distribution $l^{\star}_{\theta_{m,j}}$}

Here, the most probable value $l^{\star}_{\theta_{m,j}}$ of the log-conditional probability $l_{\theta_{m,j}}(\vec{o},\vec{\lambda})$ is derived. The latter is obtained by evaluating the value at which the rate function
$I(l_{\theta_{m,j}}/m)$ of Eq.\,(\ref{app_rate_function}) is minimized as a function of $l_{\theta_{m,j}}$ in the thermodynamic limit of $m\rightarrow\infty$. $I(l_{\theta_{m,j}}/m)$ is a positive and convex function -- see Appendix II -- and $l_{\theta_{m,j}}$ is given by a convex sum of the dynamical transition probabilities $q_{j_{{\rm B}},i,j_{{\rm A}}}$. Thus, sufficient condition for its minimization is that the identities
\begin{equation}\label{app_costraints}
\left.\frac{\partial I(l_{\theta_{m,j}}(\vec{o},\vec{\lambda})/m)}{\partial\ln q_{j_{{\rm B}},i,j_{{\rm A}}}}\right|_{l_{\theta_{m,j}}(\vec{o},\vec{\lambda}) = l^{\star}_{\theta_{m,j}}} = \,\, 0,
\end{equation}
(computed in correspondence to $l_{\theta_{m,j}}(\vec{o},\vec{\lambda}) = l^{\star}_{\theta_{m,j}}$) are all simultaneously verified for \emph{each} triplet $(j_{{\rm B}},i,j_{{\rm A}})$. If we perform the derivative of $I(l_{\theta_{m,j}}/m)$ with respect to $\ln q_{j_{{\rm B}},i,j_{{\rm A}}}$, then we find that the identities (\ref{app_costraints}) for the triplets $(j_{{\rm B}},i,j_{{\rm A}})$ apart from $(d_{\pi},d_{\lambda},d_{\pi})$ can be equivalently written by means of an unique equation, i.e.,
\begin{equation}
p_{d_{\pi},d_{\lambda},d_{\pi}}f_{j_{{\rm B}},i,j_{{\rm A}}} = p_{j_{{\rm B}},i,j_{{\rm A}}}\left(1-\widetilde{\sum_{j_{{\rm B}},i,j_{{\rm A}}}}f_{j_{{\rm B}},i,j_{{\rm A}}}\right).
\end{equation}
By summing both sides over $(j_{{\rm B}},i,j_{{\rm A}})$, we get
\begin{small}
\begin{equation}
p_{d_{\pi},d_{\lambda},d_{\pi}}\widetilde{\sum_{j_{{\rm B}},i,j_{{\rm A}}}}f_{j_{{\rm B}},i,j_{{\rm A}}} =
\left(1-\widetilde{\sum_{j_{{\rm B}},i,j_{{\rm A}}}}f_{j_{{\rm B}},i,j_{{\rm A}}}\right)\widetilde{\sum_{j_{{\rm B}},i,j_{{\rm A}}}}p_{j_{{\rm B}},i,j_{{\rm A}}},
\end{equation}
\end{small}
which, by using $\displaystyle{\sum_{j_{{\rm B}},i,j_{{\rm A}}}p_{j_{{\rm B}},i,j_{{\rm A}}}}=1$, gives
\begin{equation}\label{minimization_condition}
\widetilde{\sum_{j_{{\rm B}},i,j_{{\rm A}}}}f_{j_{{\rm B}},i,j_{{\rm A}}} = \widetilde{\sum_{j_{{\rm B}},i,j_{{\rm A}}}}p_{j_{{\rm B}},i,j_{{\rm A}}}.
\end{equation}
It is worth observing that Eq.\,(\ref{minimization_condition}) represents the condition for the minimization of the rate function $I(l_{\theta_{m,j}}(\vec{o},\vec{\lambda})/m)$ with respect to $l_{\theta_{m,j}}$. This means that the most probable trajectories of the system dynamics are those allowing for the equality between the summations of the relative frequencies $f_{j_{{\rm B}},i,j_{{\rm A}}}$ and the occurrence probabilities $p_{j_{{\rm B}},i,j_{{\rm A}}}$, respectively. Therefore, this also implies that in general the same value of $l^{\star}_{\theta_{m,j}}$ can be obtained by more than one trajectory within the configuration space of the system, each of them corresponding to a different realization of the stochastic dynamics of the system.

By combining Eq.\,(\ref{minimization_condition}) with the expression of $l_{\theta_{m,j}}(\vec{o},\vec{\lambda}) = \sum_{j_{{\rm B}},j_{{\rm A}}=1}^{d_{\pi}}\int_{\lambda}n_{j_{{\rm B}},j_{{\rm A}}}(\lambda)\ln q_{j_{{\rm B}},j_{{\rm A}}}(\lambda)d\lambda$, using Eqs.\,(\ref{app_gamma}),\,(\ref{app_f}) and substituting $l_{\theta_{m,j}}(\vec{o},\vec{\lambda})$ with $l^{\star}_{\theta_{m,j}}$, one has that
\begin{eqnarray}\label{equation_l_star_SM}
&\displaystyle{\left(\ln q_{d_{\pi},d_{\lambda},d_{\pi}} - \frac{l^{\star}_{\theta_{j,m}}}{m}\right) = (d_{\rm tot} - 1)}&\nonumber \\
&\times\displaystyle{\left(1-\widetilde{\sum_{j_{{\rm B}},i,j_{{\rm A}}}}p_{j_{{\rm B}},i,j_{{\rm A}}}\right)
\frac{p_{j_{{\rm B}},i,j_{{\rm A}}}}{p_{d_{\pi},d_{\lambda},d_{\pi}}}\gamma_{j_{{\rm B}},i,j_{{\rm A}}}}&
\end{eqnarray}
for each triplet $(j_{{\rm B}},i,j_{{\rm A}}) \neq (d_{\pi},d_{\lambda},d_{\pi})$. Then, if we extend Eq.\,(\ref{minimization_condition}) by assuming that $f_{j_{{\rm B}},i,j_{{\rm A}}} = p_{j_{{\rm B}},i,j_{{\rm A}}}$\,$\forall\,(j_{{\rm B}},i,j_{{\rm A}})$, then
\begin{equation}
\frac{p_{j_{{\rm B}},i,j_{{\rm A}}}\gamma_{j_{{\rm B}},i,j_{{\rm A}}}}{p_{d_{\pi},d_{\lambda},d_{\pi}}} = \frac{f_{j_{{\rm B}},i,j_{{\rm A}}}\gamma_{j_{{\rm B}},i,j_{{\rm A}}}}{p_{d_{\pi},d_{\lambda},d_{\pi}}}
= {\rm constant}.
\end{equation}
Since Eqs.\,(\ref{equation_l_star_SM}) have to be verified for each triplet of the system configuration space, this means that Eqs.\,(\ref{equation_l_star_SM}) are identically equivalent to the relation
\begin{eqnarray}
&\displaystyle{\frac{l^{\star}_{\theta_{m,j}}}{m} = \ln q_{d_{\pi},d_{\lambda},d_{\pi}}}& \nonumber \\
&-\displaystyle{\left(1-\widetilde{\sum_{j_{{\rm B}},i,j_{{\rm A}}}}p_{j_{{\rm B}},i,j_{{\rm A}}}\right)
\left(\widetilde{\sum_{j_{{\rm B}},i,j_{{\rm A}}}}\frac{p_{j_{{\rm B}},i,j_{{\rm A}}}}{p_{d_{\pi},d_{\lambda},d_{\pi}}}\gamma_{j_{{\rm B}},i,j_{{\rm A}}}\right)},&\nonumber \\
&&
\end{eqnarray}
finally providing us the analytical expression of the most probable distribution $l^{\star}_{\theta_{m,j}}$ as given by Eq.\,(\ref{l_star}).

\end{document}